\documentclass[twocolumn]{jpsj2}
\usepackage{revsymb}

\newcommand{\be}{\begin{equation}}\newcommand{\ee}{\end{equation}}
\newcommand{\bea}{\begin{eqnarray}}\newcommand{\eea}{\end{eqnarray}}
\newcommand{\ba}{\begin{array}}\newcommand{\ea}{\end{array}}
\newcommand{\bit}{\begin{itemize}}\newcommand{\eit}{\end{itemize}}
\newcommand{\ben}{\begin{enumerate}}\newcommand{\een}{\end{enumerate}}

\newcommand{\lab}{\label}

\newcommand{\lf}{\left}

\newcommand{\noi}{\noindent}\newcommand{\non}{\nonumber}

\newcommand{\ran}{\rangle}
\newcommand{\ri}{\right}

\newcommand{\al}{\alpha}\newcommand{\bt}{\beta}

\newcommand{\De}{\Delta}

\newcommand{\te}{\theta}

\newcommand{\om}{\omega}

\def\runtitle{Quantum Limit of Deterministic Theories}
\def\runauthor{Massimo \textsc{Blasone}, Petr \textsc{Jizba}
and Giuseppe \textsc{Vitiello}}

\title{
Quantum Limit of Deterministic Theories
}

\author{%
Massimo \textsc{Blasone},$^{1,3,}$\thanks{e-mail: m.blasone@imperial.ac.uk} Petr \textsc{Jizba}$^{2,}$\thanks{e-mail: petr@cm.ph.tsukuba.ac.jp} and Giuseppe \textsc{Vitiello}$^{3,}$\thanks{e-mail: vitiello$@$sa.infn.it} }

\inst{%
$^1$Theoretical Physics, Blackett Laboratory, Imperial College
London, SW7 2AZ London, U.K.\\
$^2$Institute of Theoretical Physics, University of Tsukuba,
Ibaraki 305-8571, Japan\\
$^3$Dipartimento di Fisica, INFM and INFN, Universit\`{a} di
Salerno, I-84100 Salerno, Italy}

\recdate{February 3, 2003}

\abst{ We show that the quantum linear harmonic oscillator can
be obtained in the large $N$ limit of a classical deterministic
system with $SU(1,1)$ dynamical symmetry. This is done in analogy
with recent work by G. 't Hooft who investigated a deterministic system based
on $SU(2)$. Among the advantages of our model based on a
non-compact group is the fact that the ground state energy is
uniquely fixed by the choice of the representation.}

\kword{ 't~Hooft's quantization, large $N$ limit, $SU(1,1)$
symmetry}

\begin{document}
\sloppy \maketitle

%%%%%%%%%%%%%%%%%%%%%%%%%%%%%%%%%%%%%%%%%%%%%%%%%%%%%%%%%%%%%%%
\section{Introduction}
%%%%%%%%%%%%%%%%%%%%%%%%%%%%%%%%%%%%%%%%%%%%%%%%%%%%%%%%%%%%%%%

Large $N$ limit, has long been recognized, plays a fundamental
r\^{o}le in obtaining  classical limit of various quantum systems.
Quantum spin models~\cite{Sim1}, quantum vector
models~\cite{Ber1} and $U(N)$ lattice gauge theories~\cite{Yaffe}
provide examples. This observation has further been reinforced by
numerical evidences~\cite{Pros1} showing that the dynamical phase
transitions in many non-integrable quantum field models in the
large $N$ limit can be identified with the stochastic transitions
from regular to chaotic motion in the corresponding classical
systems.

Yet, recently a reverse r\^{o}le of the large $N$ limit has been
conjectured by G. 't~Hooft~\cite{erice,thof1}, i.e., the large
$N$ limit of a  deterministic theory may give rise to a genuine
quantum system. This often happen in conjunction with information
loss~\cite{erice,BJV}. 't~Hooft's work has aroused substantial
interest in the possibility of obtaining a whole range of quantum
models from purely classical considerations. It should be stressed
that due to the non-local nature of the information loss and due
to emergence of (non-local) geometric phases, Bell's inequalities
cannot be utilized. The above scenario has  been studied in
numerous deterministic
systems~\cite{thof1,BJV,BCJV,carsten,Muller1}.

't~Hooft's conjecture  brings a  new perspective in the
understanding of the connection between classical chaotic dynamics
and quantum mechanics. It has been recently
shown~\cite{Muller1}, that quantum gauge field theory can
emerge in the infrared limit of a higher-dimensional, classical
(non-Abelian) gauge field theory, known to have chaotic behavior.

The purpose of this paper is to provide a concrete example
illustrating the basic features of 't~Hooft's proposal on a simple
deterministic system with a {\em non-compact} dynamical group.
Our model is alternative to the $SU(2)$ model given by 't
Hooft~\cite{erice,thof1} and presents some advantages with respect
to it in the fact that the non-compact $SU(1,1)$ structure better
fits the Weyl--Heisenberg algebra of the quantum linear harmonic
oscillator (LHO), to which our system reduces in the large $N$
limit.

In Sec.~\ref{SU2} we briefly review 't~Hooft's $SU(2)$ system. In
Sec.~\ref{SU(11)} we introduce the $SU(1,1)$ deterministic
system~\cite{BCJV} and in Sec.~\ref{QL} the corresponding quantum
limit to LHO is performed. Sec.~\ref{concl} is devoted to
concluding remarks.

%%%%%%%%%%%%%%%%%%%%%%%%%%%%%%%%%%%%%%%%%%%%%%%%%%%%%%%%%%%%%%%%
\section{'t Hooft's $SU(2)$ Model and Its Quantum Limit \label{SU2}}
%%%%%%%%%%%%%%%%%%%%%%%%%%%%%%%%%%%%%%%%%%%%%%%%%%%%%%%%%%%%%%%%

Let us briefly recapitulate 't Hooft's $SU(2)$
example~\cite{thof1}. This consists of an autonomous dynamical
system represented by $N$ distinct states $(k), \ k = 1, \ldots,
N$. The time evolution takes place in discrete time steps of equal
size, $\De t=\tau$ with periodicity condition $(k) = (k+N)$. By
admitting the following representation
\begin{equation} (0)=(N)=\lf(\ba{c} 0 \\ 0\\ \vdots \\ 1\ea\ri) \,;\,
%(1)=\lf(\ba{c} 1 \\ 0\\ \vdots \\ 0
\, \dots \,; \, (N-1)=\lf(\ba{c} 0
\\ \vdots \\ 1\\ 0\ea\ri),\\
\label{rep1}
\end{equation}
the evolution is regulated by the Hamiltonian $H$ as:
\bea U(\tau ) = e^{-i H \tau}  = e^{-i\frac{\pi}{N}}
\lf(\ba{ccccc} 0 &&&&1 \\ 1&0&&& \\ &1&0&&
\\ &&\ddots&\ddots& \\&&&1&0 \ea \ri) .
\eea
The factor $-i\pi/N$ is 't~Hooft's phase choice. Evolution matrix
satisfies the condition $U^N =-\openone$ and hence its
(un-normalized) eigenstates, say $|n\rangle, \ n = 0, 1, \ldots,
N-1$ are
\begin{equation}
|n \rangle = \sum_{k=1}^N e^{i2\pi nk /N} (k) .
\end{equation}
By defining $N\equiv 2l + 1$ and $n\equiv m+ l$ (i.e., $m = -l,
\ldots, l$) `t~Hooft mapped his system onto $SU(2)$ algebra as
follows
\bea \lab{5a}
&&\mbox{\hspace{-5mm}}H ~|l,m\ran = \omega (n+\frac{1}{2}) ~|l,m\ran . \\
&&\mbox{\hspace{-5mm}}L_3 ~|l,m\ran = m ~|l,m\ran, \non \\
&&\mbox{\hspace{-5mm}}L_+ ~|l,m\ran = \sqrt{(2l-n)(n+1)} ~|l,m+1\ran, \non \\
&&\mbox{\hspace{-5mm}}L_- ~|l,m\ran = \sqrt{(2l-n+1) n}  ~|l,
m-1\ran. \lab{5} \eea
Here $\omega = 2\pi/N\tau$. The continuous limit is obtained by
letting $l \rightarrow \infty$ and $\tau \rightarrow 0$ with
$\omega$ fixed.

\begin{figure}[h]
\hspace{1.5cm}\includegraphics[width=150pt]{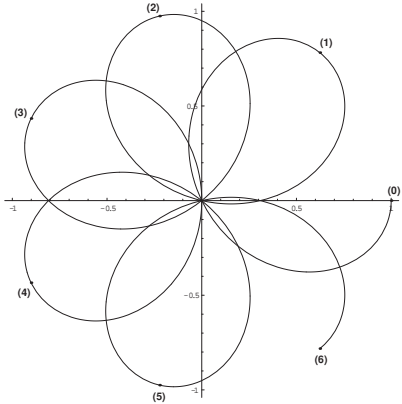}\\
 \hspace{-1.5cm} \caption{'t Hooft's
deterministic system for $N=7$. An underlying continuous dynamics
is introduced, where $x(t) = \cos (\al t) \cos(\bt t)$ and $y(t) =
- \cos (\al t) \sin(\bt t)$. At the times $t_j = j \pi/\al$, with
$j$ integer, the trajectory touches the external circle, i.e.,
$R^2(t_j)= x^2(t_j) + y^2(t_j)=1$, and thus $\pi/\al$ is the
frequency of the discrete ('t Hooft) system. At time $t_j$, the
angle of $R(t_j)$ with the positive $\mib{x}$ axis is given by: $
\te_j = j \pi - \bt t_j= j (1 - \bt/\al)\pi$. When $\bt/\al$ is a
rational number, say $q $, the system returns to the origin
(modulo $2\pi$) after $N$ steps. To ensure that the $N$ steps
cover only one circle, we have to impose $\al(t_j)= j \,2\pi/N$.
Thus, in order to reproduce 't Hooft's system for $N=7$ we choose
$q=5/7$.}\label{fig1}
\end{figure}

As shown in ref.~\citen{BCJV} such a limit corresponds to a {\em
group contraction},~and by defining $a^\dag \equiv L_+/\sqrt{2l}$
and $a \equiv L_-/\sqrt{2l}$ one recovers for $l \rightarrow
\infty $ and $\omega$ fixed the Weyl--Heisenberg algebra $h(1)$ of
quantum LHO, i.e.,
\bea
&&\mbox{\hspace{-9mm}}H~|n\ran = \omega (n+1/2)~|n\ran , \nonumber \\
&&\mbox{\hspace{-9mm}}a^\dagger ~|n\ran = \sqrt{n+1} ~|n+1\ran\;
,\; a ~|n\ran = \sqrt{n}  ~|n-1\ran  .\label{h1} \eea
't~Hooft's system may be mimicked, for instance, by a charged
particle in a cylindrical magnetron\footnote{Cylindrical magnetron
is a device with a radial, cylindrically symmetric electric field
that has in addition a perpendicular uniform magnetic field.}.
Then the particle trajectory is basically a cycloid  wrapped
around the center of the magnetron. The actual parameters (and
qualitative nature) of the cycloid are specified by the Larmor
frequency $\omega_L = qB/2m$. To implement the discrete time
evolution we confine ourself only to the observation of the largest
radius positions of the particle. So particularly we disregard any
information concerning the actual underlying trajectory. This
corresponds to loss of information. If the orbital frequency is an
integer multiple of Larmor frequency then the particle proceeds
via discrete time evolution with $\tau = 2\pi/\omega_L$ and
returns into its initial position after one revolution, see
Fig.~\ref{fig1}. The continuous limit then corresponds to an
appropriate increase in magnetic field.

%

%%%%%%%%%%%%%%%%%%%%%%%%%%%%%%%%%%%%%%%%%%%%%%%%%%%%%%%%%%%%%%%%
\section{The $SU(1,1)$ Deterministic System \label{SU(11)}}
%%%%%%%%%%%%%%%%%%%%%%%%%%%%%%%%%%%%%%%%%%%%%%%%%%%%%%%%%%%%%%%%

\begin{figure}[t]
\hspace{0cm}\includegraphics[width=250pt]{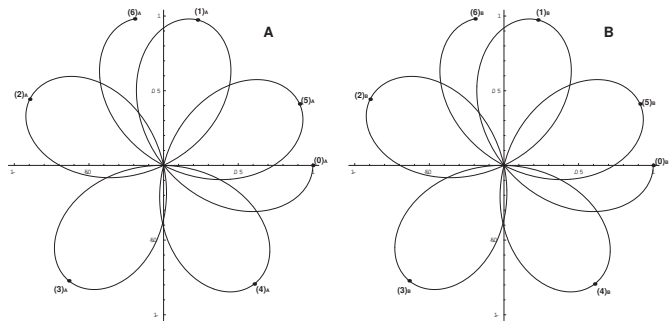}\\
 \hspace{-0cm} \caption{A
deterministic system based on $SU(1,1)$ and obtained by using the
same dynamics as in Fig.~1 with $\bt/\al = 5/3 +
\pi/40$.}\label{fig2}
\end{figure}

Let us now consider a different deterministic system, which can be
related to the $SU(1,1)$ algebraic structure. The system in
question consists of two sub-systems, each of them comprised of a
particle moving along a circle in discrete equidistant jumps. Both
particles and circle radii might be different the only common
constraint is that particles are synchronized in their jumps. Let
us further assume that for the two particles the circumference and
the elementary jump lengths are \textit{incommensurable} so that
particles never come back into the original position after a
finite number of jumps. We will label the corresponding states
(positions) as $(n)_A$ and $(n)_B$. The synchronized time
evolution materializes via discrete and identical time steps
$\triangle t = \tau$ as follows:
\begin{eqnarray*}
t \rightarrow t + \tau; &&(1)_A\rightarrow (2)_A
\rightarrow (3)_A \rightarrow (4)_A \ldots , \\
&&(1)_B\rightarrow (2)_B \rightarrow (3)_B \rightarrow (4)_B
\ldots .
\end{eqnarray*}
\noi This evolution is, of course, completely deterministic. In
fact, it is not difficult to devise a gedanken experiment
producing such an evolution. Using again the magnetron system and
assuming that the Larmor frequency and orbital frequency are
incommensurable then particles A and B ``return" into their
initial positions after infinitely revolutions, see
Fig.~\ref{fig2}.
Let us regularize the  motion by
assuming that particles come after $N$ revolutions back into
origin: $N$ is our limiting (cutoff) parameter.
Representing the actual states (positions) $(n)_{A,B}$ as in
eq.~(\ref{rep1})
then the one-time-step evolution operator acting on
$(n)_A\otimes (m)_B$ reads
\bea \non &&U(\tau ) \equiv e^{-i H \tau}  = e^{-i H_A
\tau}\otimes e^{-iH_B \tau}
\\
&&\mbox{\hspace{-9mm}}= \non \lf(\ba{cccc}  0&0&\ldots&1 \\
1&0&\ldots&0
\\ 0&1&0&\ldots\\
&&\ddots&\ddots  \ea \ri)_A \otimes \lf(\ba{cccc}  0&0&\ldots&1 \\
1&0&\ldots&0
\\ 0&1&0&\ldots\\
&\ddots&\ddots  \ea \ri)_B. \label{mat1}\eea
Note that the first lines in the matrices describe the ``spurious"
evolution
\begin{equation}
\lf(\ba{c} 0 \\ 0\\ \vdots \\ 1\ea\ri)_A\otimes \lf(\ba{c}
0\\0 \\\vdots \\ 1 \ea \ri)_B  \rightarrow \lf(\ba{c} 1\\ 0\\ \vdots \\
0\ea\ri)_A\otimes \lf(\ba{c} 1\\0\\\vdots \\ 0 \ea \ri)_B  .
\end{equation}
Such an evolution ensures that $U(\tau)$ is unitary and compatible
with the evolution on Fig.~\ref{fig2}.

 By diagonalizing $U(\tau)$ we obtain that the two matrices
involved in $U(\tau)$ have eigenvalues $\lambda_n = e^{i 2\pi
n/N}\, (n = 0,\ldots ,N-1)$ with multiplicity 1. For  future
convenience it is important to realize that
\bea \non \lambda_n = e^{i2\pi n/N}
= e^{i 2\pi \zeta n} ,
\eea%
where we have defined $\zeta = (1-N)/N$. So there is a basis
(polar basis) in which $U(\tau)$ has a  diagonal form:
\bea\label{polar}
&&U(\tau) = U_A(\tau) \otimes U_B(\tau) ,\\
\non && U_{A,B}(\tau)\equiv\mbox{diag}\left( 1, e^{i2\pi \zeta },
\ldots,  e^{i2\pi \zeta (N-1)} \right)_{A,B} . \eea
Let us denote the corresponding eigenstates of $U(\tau)$ as
$\{|n_A\rangle \otimes |m_B\rangle; n_A,m_B \in [0,(N-1)]\}$ so
that
\begin{align}\label{mat0a}
&U_A(\tau)|n_A\rangle = e^{i2\pi \zeta n_A} |n_A
\rangle ,
\\ &U_B(\tau)|m_B\rangle = e^{i2\pi \zeta m_B} |m_B \rangle  .
\label{mat0b}
\end{align}
\noi We observe that with respect to the
original non-polar basis (i.e., position vectors) the
(un-normalized) eigenvectors read
\begin{align}
&|n_A\rangle = \sum_{k=1}^{N} e^{-i 2\pi \zeta n_A k}
(k)_A , \\
& |m_B\rangle = \sum_{k=1}^{N} e^{-i 2\pi \zeta m_B k} (k)_B .
\end{align}
\noi It should be stressed that there is a deep qualitative
difference between $(\ldots)$ and $|\ldots \rangle$ vectors. The
first describe the time dependent set of states characterizing the
deterministic evolution. The second describe the time independent
states (eigenstates of the formal Hamiltonian, see below) which
have no connection with actual particle trajectory.
As a result of (\ref{mat0a}) and (\ref{mat0b}) we have
\bea |n_A\rangle \otimes |m_B\rangle = \sum_{k,l =1}^{N} e^{-i
2\pi \zeta (n_A k + m_B l)}(k)_A \otimes (l)_B  .
\label{mat01}\eea
\noi Defining $(n_A - m_B)/2 \equiv j$ (i.e., $|j|=0,1/2,1,3/2,
\ldots,$
 $(N-1)/2$ ) and $(n_A + m_B)/2 \equiv m$ (i.e., $m=|j|, |j| + 1,
\ldots, (N-1)$), we may pass from the basis $|n_A\rangle \otimes
|m_B\rangle $ to $|j,m\rangle$ basis. It is then clear that
\bea |j,m\rangle = \sum_{l,k =1}^{N} e^{-i2\pi \zeta (m (k + l) +
j(k-l))} (k)_A \otimes (l)_B  .\eea
\noi Correspondingly,
\bea \non &&U(\tau)|j,m\rangle = U_A(\tau)\otimes U_B(\tau)|j,m\rangle \\
\non &&\mbox{\hspace{-1mm}}= \sum_{k,l}^N e^{-i 2\pi \zeta (m(k+l)
+ j(k-l))}U_A(\tau)(k)_A\otimes U_B(\tau) (l)_B \\
\non &&\mbox{\hspace{-1mm}}= \sum_{k,l =1}^{N} e^{i 2\pi \zeta
(m(k+l-2) + j(k-l))}
(k)_A \otimes (l)_B\\
&&\mbox{\hspace{-1mm}}= e^{i2\pi \zeta 2 m} |j,m \rangle
.\label{mat02} \eea
\noi Similarly,
\bea \label{mat02b}
U_A(\tau)\otimes U_B(-\tau)\ |j,m\rangle=
e^{i2\pi \zeta 2 j} |j,m \rangle.
\eea
\noi We note that (\ref{mat02}) can be equivalently written as
\bea  U(\tau)|j,m\rangle = e^{i2\pi (\zeta 2m - n) } |j,m
\rangle . \label{mat03}\eea
\noi Here $n$ is an arbitrary integer allowed by the identity
$e^{i2\pi n} =1$.

%%%%%%%%%%%%%%%%%%%%%%%%%%%%%%%%%%%%%%%%%%%%%%%%%%%%%%%%%%%%%%%%
\section{Quantum Limit and Zero-Point Energy \label{QL}}
%%%%%%%%%%%%%%%%%%%%%%%%%%%%%%%%%%%%%%%%%%%%%%%%%%%%%%%%%%%%%%%%

Equation~(\ref{mat03}) implies that the total Hamiltonian has the
spectrum
\bea \non H\ |j,m\rangle &=&(H_A + H_B)\ |j,m\rangle
= \frac{2\pi}{\tau}(-  \zeta 2 m + n)\ |j,m\rangle \\
&=&\omega [(n_A + m_B) - n/\zeta]\ |j,m\rangle  ,
\eea
\noi which in the $N \rightarrow \infty$ limit is
\bea \non H\ |j,m\rangle = \frac{2\pi}{\tau}(n_A + m_B +n)\ |j,m
\rangle = \omega (2m + n)\ |j,m \rangle  . \eea
\noi From dimensionality reasons we can view  $\omega = -\zeta
2\pi/\tau$ as a formal frequency. Note that $\omega$ is finite in
the large $N$ limit; $\omega \rightarrow 2\pi/\tau$. Let us now
define:
\bea
L_3 \equiv \frac{\mathcal{H}}{\omega}  \equiv
\frac{(H_A + H_B)}{2\omega}\;
;\quad {\cal{C}} \equiv \frac{(H_A - H_B)}{2\omega}\ .
\eea
Equation~(\ref{mat02b}) implies that
\bea
 (H_A - H_B)\ |j,m\rangle = -\frac{2\pi \zeta}{\tau}\ 2j\
|j,m \rangle , \eea
\noi and  for $N\rightarrow \infty$ we have
\bea \frac{(H_A - H_B)}{2}\ |j,m \rangle = \omega j \ |j,m \rangle
 . \eea
Note that $(H_A-H_B)$ represents the energy excess of the
system A over system B and hence it is the system (dynamical)
invariant (i.e., among others, it commutes with $H$).

 Applying the
$N\rightarrow\infty$ limit and setting $n=1$ we can map our
deterministic system onto $SU(1,1)$ algebra in the following way:
\bea
L_3|j,m\rangle  = (m + 1/2)|j,m\rangle\quad ,
\quad {\cal{C}}|j,m\rangle  =  j |j,m\rangle  .
\eea
\noi  To close the algebra we need to construct the ladder
operators $L_+$ and $L_-$ in terms of the system variables. This
can be done by the following observation: We define the position
operators (or time pointers)
\bea N_{A}\ (k)_A = k(k)_A , \;\;\; ; \;\;\; N_B\ (l)_B = l
(l)_B , \eea
\noi  then
\bea &&e^{-i 2\pi \zeta (N_A + N_B)}|j,m-1\rangle  = \ |j, m \rangle  .
 \eea
\noi This in turn suggests that
\begin{eqnarray} L_+ &=& e^{-i 2\pi \zeta (N_A + N_B)}\ \left(\sqrt{\left(L_3
+ 1/2 \right)^2 - {\mathcal{C}}^2} \right)
,\nonumber \\
L_- &=& \left( \sqrt{\left(L_3 + 1/2 \right)^2 - {\mathcal{C}}^2}
\right) \ e^{i
2\pi \zeta (N_A + N_B)}\nonumber \\
&=& e^{i 2\pi \zeta (N_A + N_B)} \ \left( \sqrt{\left(L_3 - 1/2
\right)^2 - {\mathcal{C}}^2}\right),
\end{eqnarray}
or in other words positive times (positions)  ($N_{A}>0$,
$N_{B}>0$) are responsible for the forward ladder operator and
negative times (positions) ($N_{A}<0$, $N_{B}<0$) are responsible
for the backward ladder operator.
As $\cal{C}$ clearly commutes with $L_-, L_+$ and $L_3$ it
coincides with the Casimir operator of $SU(1,1)$ algebra.
Eigenvalues of the Casimir set, as usually, the algebra
representation.

Some comments are in order. Since the $SU(1,1)$ group is well
known (see e.g., ref.~\citen{Per1}), we only recall that it is locally
isomorphic to the (proper) Lorentz group in two spatial dimensions
$SO(2,1)$ and it differs from $SU(2)$ only in a sign in the
commutation relation: $[L_+, L_-] = - 2 L_3$. $SU(1,1)$
representations are well known. In particular, if we define $n
\equiv m - |j|$ and $k \equiv |j| + 1/2$ we obtain the $SU(1,1)$
discrete series $D^+_k$
\bea
&&L_3 |k,n\ran = (n+k)|k,n\ran\non, \\
&&L_+ |k,n\ran = \sqrt{(n+2k)(n+1)} \ |k,n+1\ran  , \non\\
&&L_- |k,n\ran = \sqrt{(n+2k-1)n} \ |k,n-1\ran .\label{su11}
\eea
Note that the basic parameters of $A$ and $B$ systems (such as
mass of particles or radii of circles) determine the value of
$(H_A - H_B)$ and hence the representation of $SU(1,1)$. So
particularly when both systems $A$ and $B$ are identical, then
${\mathcal{H}}$ describes the energy of the single system (be it
$A$ or $B$). The existence of the second system is then imprinted
via the choice of the representation of $SU(1,1)$. In fact, in
this particular situation we have $j =0$ which corresponds to the
fundamental representation $D^+_{1/2}$.

By identifying $|{\mbox{$\frac{1}{2}$}}, n \rangle \equiv |n
\rangle$ one has
\bea\non &&\mbox{\hspace{-7mm}} L_3 |n\ran = (n+1/2)|n\ran,
\\
&&\mbox{\hspace{-7mm}}L_+ |n\ran = (n+1) |n+1\ran\; ; \; L_-
|n\ran = n |n-1\ran . \label{ho} \eea
We note that the ladder operators matrix elements do not carry the
square roots, as on the contrary  happens in the usual
Weyl--Heisenberg algebra $ h(1)$ of LHO. In order to connect the
system eqs.~(\ref{ho}) with the quantum LHO we introduce the
following mapping in the universal enveloping algebra of
$SU(1,1)$:
\begin{equation} a= \frac{1}{\sqrt{L_3 + 1/2}}  L_- \quad ;
\quad a^\dag= L_+  \frac{1}{\sqrt{L_3 + 1/2}} ,
\label{holstein}
\end{equation}
which gives us the wanted $h(1)$ structure (\ref{h1}), with
${\mathcal{H}} =\om L_3$.
So we have found one-to-one
(non-linear) mapping between the deterministic system represented
in terms of the $SU(1,1)$ and the quantum harmonic oscillator. The
corresponding mapping (\ref{holstein}) is nothing but the
non-compact analog~\cite{gerry} of the well-known
Holstein--Primakoff representation for $SU(2)$ spin
systems~\cite{holstein}. In ref.~\citen{BCJV} this non-linear
mapping has been related to the canonical formalism for quantum
dissipation~\cite{CRV}.

We finally remark that the $1/2$ term in the $L_3$ eigenvalues is
implied by the used representation. Moreover, after a period $T =
2\pi/\omega$, the evolution of the state presents a phase $\pi$
that it is not of dynamical origin: $e^{-iHT} \neq 1$, it is a
geometric phase. Thus the zero point energy is strictly related to
this geometric phase (which confirms the result of
refs.~\citen{BJV}).
We also observe that in order to recover the
original state, one must perform one more cycle. This is related
to the isomorphism between $SO(2,1)$ and $SU(1,1)/Z_2$ and
therefore it is $e^{i 2\times 2\pi L_3} = 1$.

%%%%%%%%%%%%%%%%%%%%%%%%%%%%%%%%%%%%%%%%%%%%%%%%%%%%%%%%%%%%%%%%
\section{Conclusions\label{concl}}
%%%%%%%%%%%%%%%%%%%%%%%%%%%%%%%%%%%%%%%%%%%%%%%%%%%%%%%%%%%%%%%%

The main objective of this paper was to discuss the algebraic
structure underlying the ``quantum limit'' procedure recently
proposed by G. 't Hooft~\cite{thof1}.

 We have shown that the large
$N$ limit prescription used there for obtaining truly quantum
systems out of deterministic ones has exemplified in our case by
removing the cutoff in the regularized algebraic description of
two synchronized charged particles moving in a cylindrical
magnetron.

When the cutoff was taken to infinity we could recognize that the
algebraic underpinning of our deterministic system was that of
$SU(1,1)$. The fundamental representation $D^+_{1/2}$ then
corresponded to the situation when the second particle was
``forgotten". In this case we were able to find a one-to-one
mapping of our $SU(1,1)$ deterministic system onto the quantum
LHO\@. Such a mapping is an analog of the well known
Holstein--Primakoff mapping for diagonalizing  the ferromagnet
Hamiltonian~\cite{holstein}.

%%%%%%%%%%%%%%%%%%%%%%%%%%%%%%%%%%%%%%%%%%%%%%%%%%%%%%%%%%%%%%%%
\section*{Acknowledgments}
%%%%%%%%%%%%%%%%%%%%%%%%%%%%%%%%%%%%%%%%%%%%%%%%%%%%%%%%%%%%%%%%

%We would like to thank to organizers of the Waseda International
%Symposium on Fundamental Physics - ``New Perspectives in Quantum
%Physics" for the simulating environment.
The authors thank the ESF network COSLAB, MIUR, INFN, INFM, EPSRC and
JSPS for partial support.

\end{document}